\documentclass[aps,pra,twocolumn,superscriptaddress,showpacs,showkeys,amsmath,amssymb]{revtex4}

\usepackage{amsfonts}
\usepackage{amssymb,amsmath}
\usepackage{mathrsfs}
\usepackage{latexsym}
\usepackage{amsmath}
\usepackage[cp1251]{inputenc}
\usepackage{graphicx}
\usepackage{dcolumn}
\usepackage{bm}
\usepackage{color}

\RequirePackage{ifthen}
\RequirePackage[pdfstartview=FitH]{hyperref}
\begin{document}
	
	\title{Bipolaron in one-dimensional $SU(3)$ fermions\\ with three-body interaction}
	\author{O.~Hryhorchak}
	\affiliation{Professor Ivan Vakarchuk Department for Theoretical Physics, Ivan Franko National University of Lviv, 12 Drahomanov Street, Lviv, Ukraine}
	\author{G.~Panochko} \affiliation{Department of Optoelectronics and Information Technologies, Ivan Franko National University of Lviv, 107 Tarnavskyj Str., Lviv, Ukraine}
	\author{V.~Pastukhov\footnote{e-mail: volodyapastukhov@gmail.com}}
	\affiliation{Professor Ivan Vakarchuk Department for Theoretical Physics, Ivan Franko National University of Lviv, 12 Drahomanov Street, Lviv, Ukraine}

	\date{\today}

	
	
	\begin{abstract}
		The properties of the one-dimensional $SU(3)$ population-imbalanced fermions are discussed. The system is assumed to be in the two-body resonance where all two-body scattering lengths diverge, and the only interaction between fermions that is taken into account is the short-range three-body one. In particular, we consider the situation when only one `flavor' of fermions is macroscopically occupied, and there are exactly two atoms of two others. This system supports the trimer and the medium-induced dimer states studied here in detail and shows evidence of color superfluidity.
	\end{abstract}
	
	\maketitle
	\section{Introduction}
	In recent years, the one-dimensional systems of bosons and fermions with a contact three-body potential, typically appearing as the confinement-induced effective forces \cite{Mazets_08} or as a residual interaction \cite{Valiente_19,Pricoupenko_Petrov_19,Pricoupenko_Petrov_21} between three atoms with divergent two-body scattering lengths, attracted much attention in the literature. Being efficiently treated in a few-body limit \cite{Sekino_18,Nishida_18,Guijarro_et_al,Pricoupenko_18,McKenney_19}, the $\delta$-like three-body (pseudo)potential is the simplest and most relevant integrability-breaking interaction for the macroscopic systems. The renormalizability of the three-body interaction in one dimension \cite{Drut_18} allows an analysis of the many-body properties for very dilute systems \cite{Pastukhov_19} relying on the perturbative consideration, or at the high-temperature region \cite{Maki_19,McKenney_20} where the virial expansion even for the harmonically-trapped quantum gases \cite{Czejdo_20} is possible. An approximately obtained many-body behavior is strongly affected \cite{Valiente_Pastukhov} by an intrinsic, for the one-dimensional systems with three-body interaction, quantum scale anomaly \cite{Daza_19}. In combination with weak two-body attraction, the repulsive three-body potential can stabilize \cite{Morera_22} the quantum droplet and liquid states in gaseous systems. It is also mandatory \cite{Sekino_Nishida_21} for the correct effective field theory description of the Bose-Fermi duality \cite{Valiente_21,Valiente_20} in one dimension. In higher fractional dimensions below $d=2$, the effective field theory description of these systems can be formulated \cite{Hryhorchak_22} by successive renormalization of the three-, four- and higher-body sectors or by generalization to the two-channel model \cite{Polkanov_24}. 
	
	Generally, the thermodynamic limit of one-dimensional systems with the three-body interaction is not well-studied. Even the detailed ground-state phase diagram of a macroscopic number of the $SU(3)$ fermions is still challenging, and the in-medium three-body states were only explored \cite{Tajima_22} within the three-particle ladder approximation. It contrasts the systems of the same symmetry group and with the two-body interaction in three- \cite{Paananen_06,He_06,Catelani_08,Floerchinger_09,Ozawa_10,Nummi_11}, and two-dimensional \cite{DeSilva_09,Salasnich_11,Kirk_17} geometries much. Being realized in experiments \cite{Ottenstein_08} with $^6$Li atoms in three different hyperfine states, the $SU(3)$ fermionic mixture on the mean-field level represents two paired components and one unpaired. However, this is only true for weakly attracting systems, and the phase diagram of the three-component fermions, in general, should include regions with trimers and mixtures of superfluids. This leads to a physical picture somewhat resembling the quark matter \cite{Rapp_07,Nishida_12,Tajima_19}. Recently, this interplay between color superfluidity and the trimers' formation on the one-dimension lattice was addressed \cite{Chetcuti_23} within the combination of Bethe ansatz and numerical methods. At finite densities, another tripling mechanism can occur \cite{Akagami_21,Tajima_21}, somewhat similar to the Cooper pairing but involving three fermions.
	
	The present article is a step forward in understanding the one-dimensional $SU(3)$ Fermi gases with three-body interaction in the thermodynamic limit that deals with the properties of a bipolaron, i.e. one macroscopically occupied component with exactly two atoms of another `flavors' immersed.

	\section{Formulation}
	\subsection{Model and method description}
	We consider a model of three-component fermions of mass $m$ loaded in one-dimensional volume $L$ with periodic boundary conditions. The system is assumed to be fine-tuned to the two-body resonance where the two-body scattering lengths diverge, and the only interaction between fermions considered is the short-range three-body one. The Euclidean action describing this model is the following
	\begin{eqnarray}\label{S}
	\mathcal{A}=\int dx \sum_{\sigma=1,2,3}\psi^{\dagger}_{\sigma}\left\{\partial_{\tau}-\xi_{\sigma}
	\right\}\psi_{\sigma}\nonumber\\
	-g_{3,\Lambda}\int dx\,n_1n_2n_3,
	\end{eqnarray}
	where $x=(\tau, z)$ with $\tau\in [0,\beta)$ (limit $\beta \to \infty$ is taken in final formulas), $z\in [0,L)$, and $\xi_{\sigma}=\varepsilon-\mu_{\sigma}=-\frac{1}{2m}\partial_{z}^2-\mu_{\sigma}$ ($\hbar=1$) with $\mu_{\sigma}$ being the chemical potentials that fix the total number of particles in each component. We also introduced notations for antiperiodic in imaginary time $\tau$ complex Grassmann fields $\psi^{\dagger}_{\sigma}(x)$, $\psi_{\sigma}(x)$ and for local densities of each fermionic sort $n_{\sigma}(x)=\psi^{\dagger}_{\sigma}(x)\psi_{\sigma}(x)$. The bare coupling constant $g_{3,\Lambda}<0$, characterizing the depth of delta-like attractive three-body potential well with the vacuum three-particle bound-state energy \cite{Guijarro_et_al,Valiente_19} $\epsilon_3=-\frac{4e^{-2\gamma}}{ma^2_3}$ ($\gamma=0.5772\ldots$ is the Euler-Mascheroni constant), can be explicitly written in momentum space \cite{Drut_18,Pastukhov_19}
	\begin{eqnarray}\label{g_3}
	-g^{-1}_{3,\Lambda}=\frac{1}{L^2}\sum_{p, p'}\frac{1}{\varepsilon_p+\varepsilon_{p'}+\varepsilon_{p+p'}+|\epsilon_3|},
	\end{eqnarray}
	($\varepsilon_p=\frac{p^2}{2m}$ and ultraviolet cutoff $\Lambda$ is implied on the wave-vector summations). In the following, we assume the minimal many-body composition for the three-body interaction to be nonzero, namely, only one component of fermions is macroscopically occupied (say, component `3') with density $n$ and there are exactly two atoms of another species (one of each sort `1' and `2'). In the thermodynamic limit, the properties of host fermions (sort `3') cannot change drastically when a few impurities are immersed. Therefore, both the spectrum and the Fermi step-function-like momentum distribution of particles of sort `3' only gain corrections of order $1/L$. The same actually concerns the single-particle energy dispersion of impurities. Two latter facts simplify the whole field theoretical analysis further, which should be held without renormalizing the single-particle propagator of host fermions and impurities. To distinguish the trimer and dimer phases of the system, we perform the calculations of the appropriate three- and two-particle propagators, respectively. In practice, however, one has to compute the three-to-three and two-to-two vertices. Their poles determine the energy of two impurities in the trimer and dimer states separately. Then, the phase that realizes the global energy minimum should be identified with the thermodynamically stable one.
	
	\subsection{Single particle-hole approximation}
\begin{figure}[h!]
		\centerline{\includegraphics
			[width=0.45
			\textwidth,clip,angle=-0]{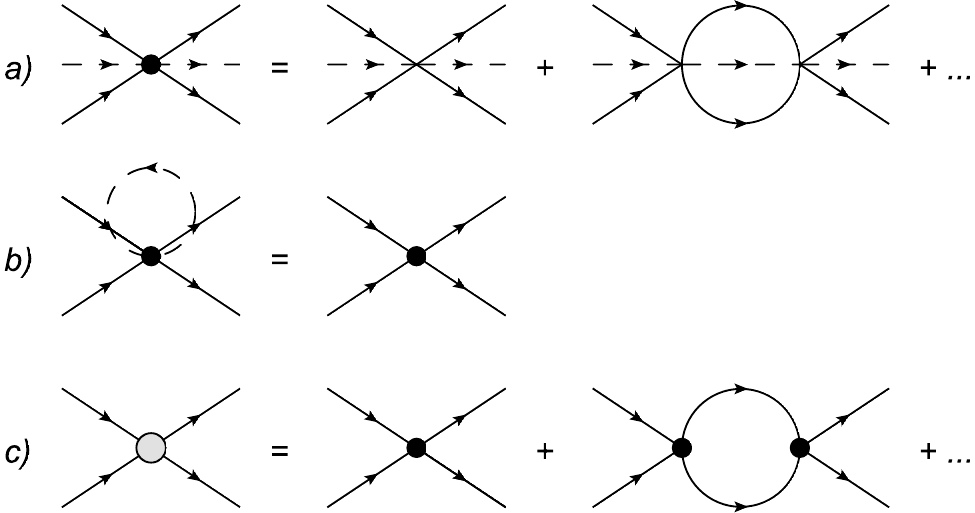}}
		\caption{Diagrams standing for various vertices: a) the trimer propagator $\mathcal{T}_{3}(P)$; b) the induced two-body coupling (\ref{g_2}); c) the two-impurity vertex $\mathcal{T}_{2}(S)$. Solid and dashed lines represent the impurity (sort `1' and `2') and the host fermion (sort `3') propagators, respectively.}\label{vertices_fig}
	\end{figure}

	The properties of two impurities are well-understood in the limit of extreme diluteness $na_3\ll 1$ of the considered system. In this case, the fingerprints of the vacuum three-body bound state are still tangible. Therefore, in the region of small gas parameters, two impurities, `1' and `2', form the trimer, which additionally involves one atom from the medium. The properties of such a trimer state in dilute limit are encoded in the simplest series of diagrams represented in Fig~\ref{vertices_fig}(a).
	Fortunately, the above diagrammatic series sums up to the very end in the $(1+1)$ Fourier space with the result (here $P$ is the total two-momentum of three particles)
	\begin{eqnarray}\label{Tau_3}
	\mathcal{T}^{-1}_{3}(P)=g^{-1}_{3,\Lambda}+\frac{1}{L}\sum_{k}\Pi_i(P+K)|_{i\nu_k\to -\xi_k},\\
	\Pi_i(S)=\frac{1}{L}\sum_{p}\frac{1}{2\varepsilon_p-2\mu_i+\varepsilon_s/2-i\omega_s},
	\end{eqnarray}
	that can be written in terms of elementary functions (from now on $\xi_k=\varepsilon_k-\mu$, where $\mu=\mu_3=\frac{p^2_F}{2m}$ is the Fermi energy of one-dimensional ideal fermion gas of density $n$ and $\mu_1=\mu_2=\mu_i$; additionally $|k|>p_F$ and $|q|<p_F$). The cutoff dependence of sum over $k$ in a vertex function, $\mathcal{T}_{3}(P)$, exactly cancels out with the $\Lambda$-dependent term in the inverse bare coupling $g^{-1}_{3,\Lambda}$. Reminding that the vertex (\ref{Tau_3}) contains all information about two impurities in the trimer state is superfluous.

	An important consequence of the macroscopic occupation of component `3' is the emergence of the induced two-body interaction between impurities. In the simplest case (dilute limit), this effective potential $g_2(S)$, which depends on the total energy and the total momentum of colliding `1' and `2' particles, is given by the diagram in Fig.~\ref{vertices_fig}(b)
	\begin{eqnarray}\label{g_2}
	g_{2}(S)=\frac{1}{L}\sum_{q}\mathcal{T}_{3}(S+Q)|_{i\nu_q\to \xi_q}.
	\end{eqnarray}
	The physics of the system crucially depends on the nature of the effective potential, i.e. sign of the coupling $g_{2}(S)$ at low energies of colliding impurities. The induced two-impurity potential is strictly repulsive at very low densities $n$ of host fermions. However, increasing the density of sort `3' particles leads to changing its character to an attractive one. Since the trimer interacts repulsively with 3' particles, we simultaneously lower the trimer binding energy by making the system more dense. Therefore, at some critical magnitude of $a_3p_F$, dimer formation should be more energetically preferable, and the transition from trimer to dimer state of two impurities should be observed. In order to study a potential emergence of this phenomenon in the adopted approximation, we construct a type of the Bethe-Salpeter equation for the two-body vertex function $\mathcal{T}_{2}(S)$ by summing up series in Fig.~\ref{vertices_fig}(c). Actually, this is an analogue of the Schr\"odinger equation (with the degree of freedom that describes a relative motion of two atoms has been integrated out) for the center-of-mass motion of two impurities that mutually interact through the induced (energy-dependent) potential $g_2(S)$. The function $\mathcal{T}_{2}(S)$ predetermines the possible two-body bound states of impurities as well as the information about their scattering properties. The summation of diagrams in Fig.~\ref{vertices_fig}(c) leads to the result
	\begin{eqnarray}\label{Tau_2}
	\mathcal{T}^{-1}_{2}(S)=g^{-1}_{2}(S)+\Pi_i(S).
	\end{eqnarray}
	A vertex function $\mathcal{T}_{2}(S)$ is straightforwardly related to the two-impurity propagator $G_2(S)=-\Pi_i(S) +\mathcal{T}_{2}(S)\Pi^2_i(S)$
	\begin{eqnarray}\label{G_2}
	G_2(S)=-\frac{1}{\Pi^{-1}_i(S)+g_{2}(S)}
	\end{eqnarray}
	[where $-\Pi_i(S)$ and $g_{2}(S)$ should be treated as a free propagator and the simplest self-energy insertion, respectively] that contains complete information about the dimer state in the adopted approximation.

	\subsection{Results}
\begin{figure}[h!]
		\centerline{\includegraphics
			[width=0.245
			\textwidth,clip,angle=-0]{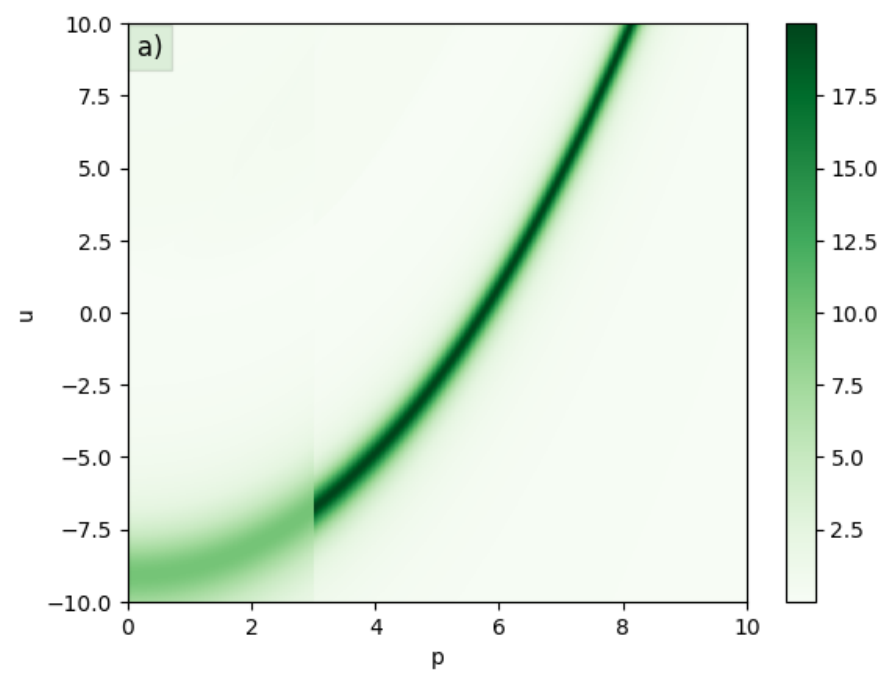}
			\includegraphics
			[width=0.235
			\textwidth,clip,angle=-0]{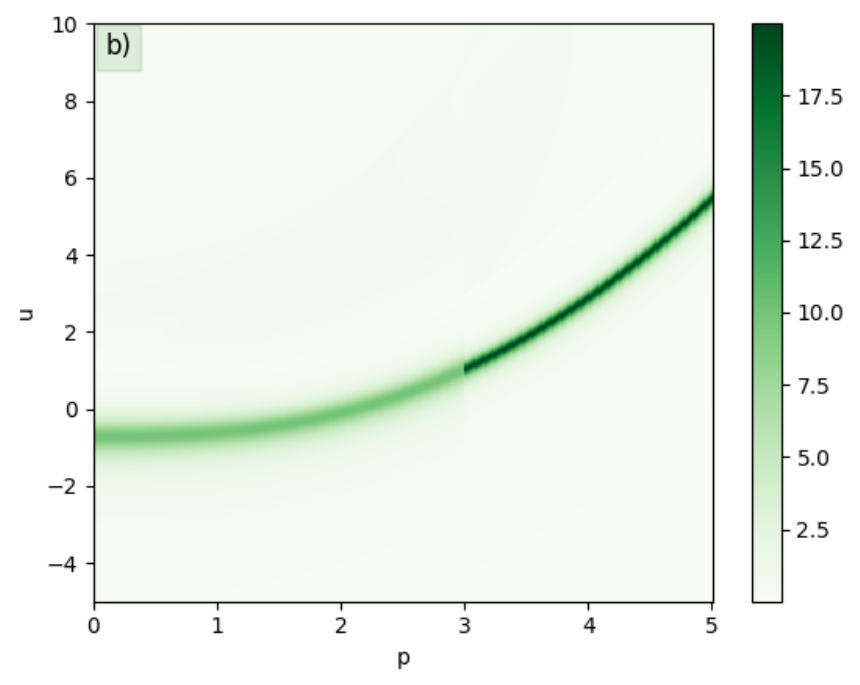}}
		\centerline{\includegraphics
			[width=0.23
			\textwidth,clip,angle=-0]{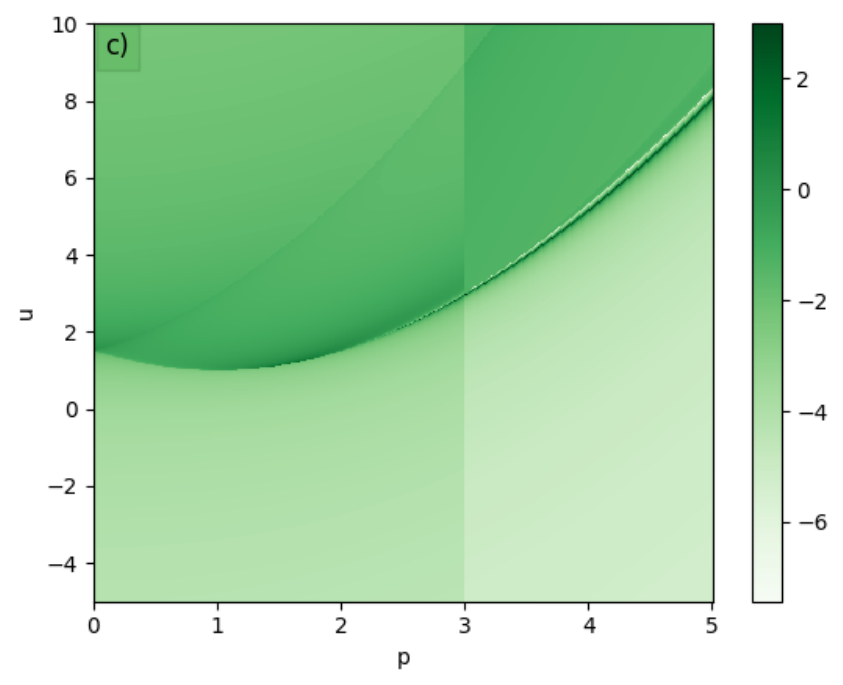}
			\includegraphics
			[width=0.23
			\textwidth,clip,angle=-0]{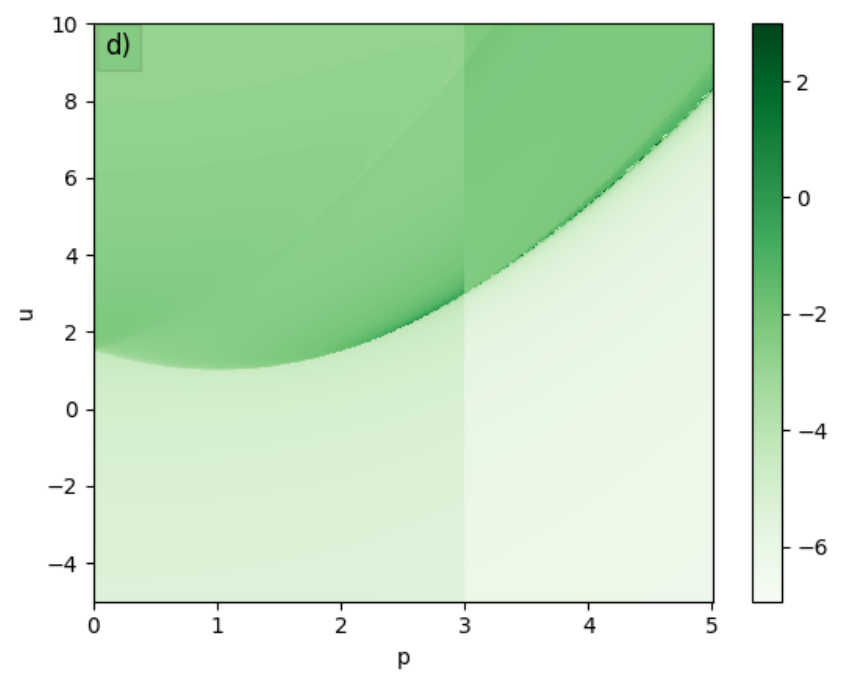}}
		\caption{Spectral density of the trimer at various coupling strengths: $\ln(|\epsilon_3|/\mu)=3$ (a), $\ln(|\epsilon_3|/\mu)=2$ (b), $\ln(|\epsilon_3|/\mu)=0$ (c), and $\ln(|\epsilon_3|/\mu)=-2$ (d) (the last two in logarithmic scale). Parameters $u$ and $p$ are energy (in units of $\mu$) and momentum (in units of $p_F$), respectively. We used the peak broadening $0.05\mu$.}\label{trimer_fig}
	\end{figure}

	Let us briefly discuss the trimer state of the system. The requirement of the microscopical population of the trimer state, $\mathcal{T}^{-1}_{3}(P)|_{i\nu_p\to i0^+, p=p_0}=0$, (where $p_0$ realizes minimum of the trimer dispersion) fixes the impurity chemical potential $\mu_i$. Considering vacuum-like trimer ($p_0=0$), one can obtain $\mathcal{T}_{3}(P)=Z_3(p)\left[i\nu_p-\frac{p^2}{2M^*_3}\right]^{-1}$ near its pole at small $p$, the propagator identifies the trimer parabolic dispersion relation with the effective mass $M^*_3$. The latter can be calculated exactly $\frac{3m}{M^*_3}=1-\frac{\sqrt{2|\epsilon_3|/3\mu}}{2(\sqrt{2|\epsilon_3|/3\mu}-1)^2}$ as well as the binding energy of motionless trimer $\frac{1}{2}-\frac{3}{2}(\sqrt{2|\epsilon_3|/3\mu}-1)^2$ (in units of $\mu$). In the dilute limit ($|\epsilon_3|/\mu\gg 1$), the trimer effective mass $M^*_3/(3m)=1+\frac{\sqrt{3}e^{\gamma}}{8}a_3p_F+\ldots$ is only slightly larger than triple mass of atom. The residue $Z_3(0)/Z^{\textrm{vac}}_3(0)=1-\sqrt{3\mu/(2|\epsilon_3|)}$ (in comparison to its vacuum magnitude) at low momenta reveals the breakdown of the quasiparticle picture description at $a_3p_F\ge\frac{4e^{-\gamma}}{\sqrt{3}}\simeq 1.3$.  The intense interaction with the fermions completely destroys the vacuum-like trimer state in the strong-coupling limit $|\epsilon_3|/\mu\le6$ providing the total binding energy of two impurities is not enough to catch the host fermion, i.e. $2\mu_i+\mu\ge0$. Below this value, $|\epsilon_3|/\mu=6$, the effective mass is negative, signalling the shift of minimum of the trimer dispersion towards non-zero momentum $p_0$. A very similar situation is meat \cite{Parish2013} in the molecule behavior of the two-dimensional Fermi polaron with pairwise interaction. We note that formally for $|\epsilon_3|/\mu\le 3/8$, the effective mass $M^*_3$ is again positive definite, but the trimer itself does not exist. At large momenta, the dressed trimer dispersion relation asymptotically touches its vacuum counterpart $\epsilon_3+\frac{p^2}{6m}$. Of course, the most complete information about the fate of impurities in this phase is represented by the trimer spectral density $\Im\mathcal{T}_{3}(P)|_{i\nu_p\to \nu+i0^+}$ (see Fig.~\ref{trimer_fig}). 
	Particularly, these figures confirm the above analysis. At large $|\epsilon_3|/\mu$, as it results from Fig.~\ref{trimer_fig}(a),(b), one observes the well-defined trimer, while at strong couplings it disappears in the three-body continuum Fig.~\ref{trimer_fig}(c),(d) (note that here we used log-scale for better visualization).

	We have also explored the properties of the dimer state. Particularly, the binding energy (the doubled impurity chemical potential $2\mu_i$) of two impurities is determined by poles of $G_2(S)$ at zero $(1+1)$ momentum. In the weak-coupling limit $a_3p_F\ll 1$, we find $2\mu_i=\epsilon_3-\frac{2}{9}\mu+\dots$, while the inverse one is characterized by the logarithmically slow $2\mu_i/\mu=-6/\left(\ln\frac{\mu}{|\epsilon_3|}\right)^2+\dots$ tending to zero with the increasing of $\epsilon_3$. The dimer dispersion at small $s/p_F$ shifts upward by a quadratic correction $\frac{s^2}{2M^*_2}$, with $M^*_2$ being the dimer effective mass. 
	\begin{figure}[h!]
		\centerline{\includegraphics
			[width=0.6
			\textwidth,clip,angle=-0]{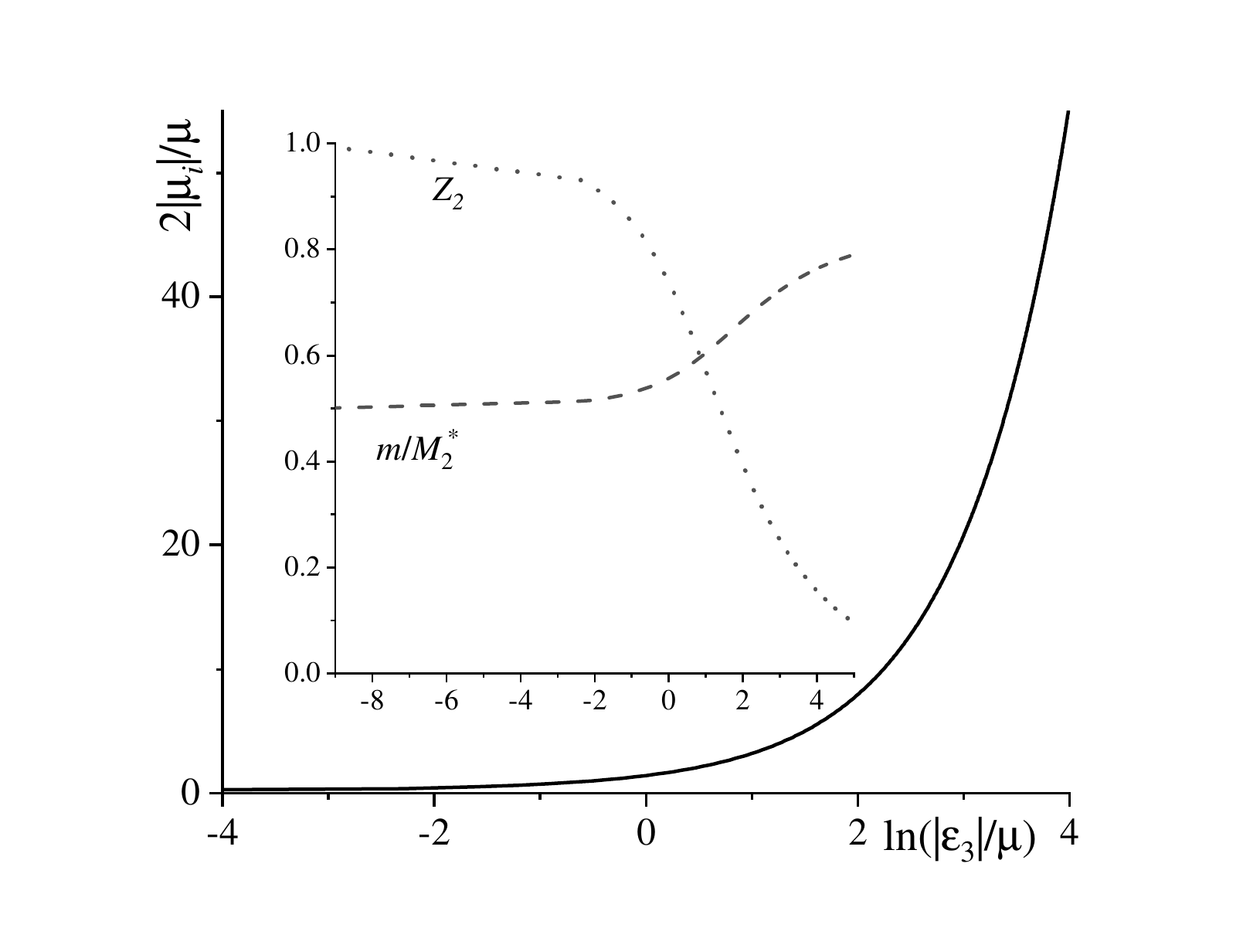}}
		\caption{Parameters of the dimer low-momentum state: absolute value of the binding energy (solid line), inverse effective mass (dashed line) and rescaled quasiparticle residue (dotted line).}\label{dimer_fig}
	\end{figure}
	In Fig.~\ref{dimer_fig}, we present the results of numerical calculations for the low-momentum parameters of the dimer spectrum. Figure~\ref{dimer_fig} also displays the residue of the dimer propagator near the pole at zero momentum $Z_2(0)/Z^{\textrm{vac}}_2(0)$ (rescaled to its vacuum magnitude). It is an important characteristic of the emergence of the dimer state in the system. Particularly, we see that at high densities $a_3p_F\gg 1$, the dimer state is well-defined while being strongly suppressed at low densities of host fermions. The obtained dimer state generally manifests qualitatively similar behavior to that of the attractive branch \cite{Schmidt_etal} of a standard two-dimensional Fermi polaron.

	\section{Summary and discussion}
	Let us summarize the obtained phase diagram of two non-identical impurities interacting through the three-body contact potential with the macroscopic number of fermions in one dimension. At small densities of host fermions, our $\mathcal{T}$-matrix calculations revealed the well-defined trimer state of the system with a slightly modified dispersion relation. The increase of the medium density leads to suppression of the trimer state, and the latter completely vanishes at some critical magnitude of interaction $\ln(|\epsilon_3|/\mu)=\ln 6\approx 1.79$. On the other side of the phase diagram, where the density of fermions is extremely large $\ln(\mu/|\epsilon_3|)\ll 1$, the attachment of a single host atom to form a trimer costs an enormous amount of energy, and the dimer (bipolaron) with the logarithmically vanishing binding energy emerges. This well-defined quasiparticle (see $Z_2$ in Fig.~\ref{dimer_fig}) that cannot occur in a vacuum possesses an almost Galilean-invariant spectrum with the total mass of two impurities. The decrease of density of the medium fermions almost totally depletes the dimer as a quasiparticle at $\ln(|\epsilon_3|/\mu)\approx 2\div 4$. Therefore, one may expect the dimer-trimer transition in this region. It turns out, however, that the dimer state is {\it always} energetically more preferable over the trimer state, and no transition happens. Even at very low densities of host fermions up to the one-atom limit, the binding energy of dimer is lower according to our calculations.
	Similarly to the conventional Fermi polaron problem, the adopted $\mathcal{T}$-matrix approximation does not treat the trimer and dimer on equal footing \cite{Combescot_08,Mathy_11}. One may suggest that a more sophisticated treatment lowers the trimer energy. However, there is some reasoning \cite{Hryhorchak_24} based on the ansatz for trimer state, which captures the particle-hole excitations appropriately that the dimer-trimer transition never occurs in the one-dimensional $SU(3)$ fermions with contact three-body interaction at extreme population imbalance of one `flavor'. Despite the latter fact, our findings -- the obtained quasiparticle residues -- give confidence in the experimental preparation of both long-lived metastable trimers and dimers in the system's low- and high-density limits, respectively. Similarly to fermions \cite{Tononi_22} with two-body contact interaction, the stable trimers, tetramers, etc., should emerge in the extension of the considered model to a strong imbalance between atom masses of different `flavors'. Finally, our findings give the first evidence for potential color superfluidity in the one-dimensional $SU(3)$ fermions with three-body interaction at finite densities of all constituents.

	\begin{center}
		{\bf Acknowledgements}
	\end{center}
	We thank Dr. I. Pastukhova for carefully reading the manuscript. Project No.~0122U001514 partly supported this work from the Ministry of Education and Science of Ukraine.

	\section{Appendix}
	For completeness, here we present an explicit analytical expression for the three-body $\mathcal{T}$-matrix (sum of diagrams in Fig.~\ref{vertices_fig} (a)). Taking into account the result for the two-impurity bubble
	\begin{eqnarray*}
		\Pi_i(S)=\frac{\sqrt{m}}{2}\frac{1}{\sqrt{\varepsilon_s/2-2\mu_i-i\omega_s}},
	\end{eqnarray*}
	it is straightforward to calculate the trimer propagator (\ref{Tau_3})
	\begin{eqnarray*}
		&&\frac{2\sqrt{3}\pi}{m}\mathcal{T}^{-1}_{3}(P)=\ln(2|\epsilon_3|/3\mu)-f(\bar{p},i\bar{\nu}_p)\\
		&&-\textrm {sg}(1-\bar{p}/3)f(-\bar{p},i\bar{\nu}_p)
		-\Theta(\bar{p}/3-1)\ln\left[u(\bar{p},i\bar{\nu}_p)\right],
	\end{eqnarray*}
	where $\bar{p}=p/k_F>0$, $\bar{\nu}_p=\nu_p/\mu$ and $u(\bar{p},i\bar{\nu}_p)=\bar{p}^2/3-1-2\mu_i/\mu-i\bar{\nu}_p$. $\textrm {sg}(x)$ and $\Theta(x)$ are standard sign and the Heaviside functions, respectively. Additionally, we introduced $f(\pm\bar{p},i\bar{\nu}_p)=\ln\left(|1\pm\bar{p}/3|+\sqrt{(1\pm\bar{p}/3)^2+2u(\bar{p},i\bar{\nu}_p)/3}\right)$. Parameters of the dimer spectrum cannot be represented via elementary functions and were calculated by numerical integration.

\end{document}